\documentclass[12pt]{article}

\usepackage{times}
\usepackage{amsmath}
\usepackage{amssymb}
\usepackage{amsfonts}
\usepackage{epsfig}

\def\invstar{{\star^{-1}}}

\def\pd#1{{\partial_{#1}}}
\def\d{{\rm d}}
\def\C{\mathcal{C}}
\def\M{\mathcal{M}}
\def\O{\mathcal{O}}
\def\D{\mathcal{D}}

\def\AA{\mathcal{A}}
\def\FF{\mathcal{F}}
\def\GG{\mathcal{G}}
\def\TT{\mathcal{T}}

\def\ff{f}

\def\i{\imath}

\def\in{\textrm{in}}
\def\out{\textrm{out}}
\def\g{\textrm{g}}
\def\e{\textrm{em}}

\def\u{\mu}
\def\v{\nu}

\def\U{\hat{\mu}}
\def\V{\hat{\nu}}
\def\X{\hat{x}}
\def\Y{\hat{y}}
\def\T{\hat{t}}
\def\Z{\hat{z}}

\def\be{\begin{eqnarray}}
\def\ee{\end{eqnarray}}

\begin{document}
\vspace{0.2cm} {\Large \centerline{{\bf
Scattering of Spinning Test Particles by}} \vskip 8pt \centerline{{\bf
Plane Gravitational and Electromagnetic Waves}} \vskip 30pt}

{\large \centerline
{{S Kessari, D Singh, R W Tucker, and C Wang}}
\vskip 5pt
\centerline{Department of Physics, Lancaster University, Lancaster LA1
4YB, UK} }
\vskip 8pt
\centerline{February 2002} \vskip 30pt

\abstract {The Mathisson-Papapetrou-Dixon (MPD) equations for the
motion of electrically neutral massive spinning particles are
analysed, in the pole-dipole approximation, in an Einstein-Maxwell
plane-wave background spacetime.  By exploiting the high symmetry
of such spacetimes these equations are reduced to a system of
tractable ordinary differential equations.  Classes of exact
solutions are given, corresponding to particular initial
conditions for the directions of the particle spin relative to the
direction of the propagating background fields.  For
Einstein-Maxwell pulses a scattering cross section is defined
that reduces in certain limits to those associated with the
scattering of scalar and Dirac particles based on classical and
quantum field theoretic techniques.  The relative simplicity of
the MPD approach and its use of macroscopic spin distributions
suggests that it may have advantages in those astrophysical
situations that involve strong classical gravitational and
electromagnetic environments.}

\section{Introduction}

Einstein's theory of gravitation predicts that certain
non-stationary distributions of matter will produce dynamic
changes in the geometry of spacetime. Those that propagate are
known as gravitational waves. Disturbances that also involve
accelerating sources of the electromagnetic field are also
expected to produce electromagnetic waves. Such scenarios can
occur in many violent astrophysical processes and the energetics
of such processes are poorly understood in many cases. It has also
been suggested that matter with spin may play a dual role in many
astrophysical processes that emit electromagnetic and
gravitational waves.  The constituents of plasmas that participate
in the generation of such waves may become polarised in strong
magnetic fields and spinning matter may respond to tidal forces
that include interactions of spacetime curvature with angular
momentum. Such processes belong to the realm of general
relativistic transport theory. However some insights into the
interaction of spinning test matter with gravitational and
electromagnetic waves can be gained by studying the classical
motion of spinning test particles in plane gravitational and
electromagnetic field solutions to the Einstein-Maxwell equations.
In this paper we explore the motion of massive  electrically
neutral particles with spin in such fields. Since the
electromagnetic field influences the spacetime metric it will have
an effect on such electrically neutral particles even in the
absence of gravitational waves.

 By neglecting self-gravitation and back-reaction the dynamics of
classical test particles with angular momentum was first studied
in detail by Mathisson, Fock, Papapetrou {\it et al}
\cite{papapetrou}. The theory was further clarified by Dixon
\cite{dixon1,dixon} using a rationalised multipole expansion technique
and developed by Ehlers, Rudolph \cite{ehlers} and others. In this
article we employ the pole-dipole approximation so that the motion
of a classical spinning test particles is governed by the
Mathisson-Papapetrou-Dixon (MPD) equations. Such MPD equations
predict gravitational spin-spin interactions between rotating
stars and orbiting massive spinning particles.

 The properties of plane
 gravitational  wave spacetimes  have attracted considerable attention due to
 their
 high symmetry~\cite{penrose}.
This has led to  investigations of colliding gravitational and
electromagnetic waves~\cite{griffiths} and the scattering of
massive  particles in such backgrounds~\cite{ferrari}. The
scattering of  scalar fields was considered by
Gibbons~\cite{gibbons}, Garriga and Verdaguer~\cite{garriga} and
others. The corresponding differential scattering cross-sections
were found to be similar to those for the scattering of  classical
spinless particles. The scattering of electrically neutral Dirac
fields was subsequently investigated by Bini and
Ferrari~\cite{bini_ferrari} and compared with
 that involving classical and quantum spinless  particle scattering.

 The classical
scattering of  neutral {\it spinning} particles by plane
gravitational waves   has been considered from a number of
different perspectives \cite{nieto, bini, mohseni}. In a recent
paper~\cite{mtw} a special class of solutions of the MPD equations
describing the non-geodesic motion of a massive spinning test
particle in plane gravitational wave spacetime was constructed.
The spin-curvature coupling was shown to give  rise to parametric
excitations of
 spinning matter by harmonic plane gravitational waves.

In this paper we generalise this analysis and suggest that
information about the scattering of electrically neutral spinning
matter in an Einstein-Maxwell plane wave background can be gleaned
from solutions to the MPD equations by exploiting the high Killing
symmetry of such spacetimes.

In section (2) the basic MPD equations are summarised and an
expression given for constants of the motion that can be generated
from Killing symmetries. Such expressions are pivotal in the
subsequent analysis. In section (3) the MPD equations are
expressed in different charts adapted to Einstein-Maxwell plane
wave spacetimes. Such charts facilitate the the symmetry reduction
of the equations to a tractable system of ordinary differential
equations. In section (4) several exact solutions to these
equations are discussed. The final section exploits the properties
of an Einstein-Maxwell pulse (sandwich spacetime) in order to
estimate  the scattering of an arbitrary  distribution of
classical spins according to the MPD equations. The result is
compared with the classical and quantum scattering cross-sections
of scalar and spinning particles calculated by other techniques in
such backgrounds.

\section{Equations of motion for neutral spinning test particles}

In the monopole-dipole approximation the MPD equations determine
the world-line of a spinning test particle with tangent vector
$V$, momentum vector $P$ and spin 2-form $s$ in terms of the
spacetime metric $g$, its Levi-Civita connection $\nabla$ and
Riemannian curvature tensor $R$. \footnote{Units in which $c = G =
1$ are used throughout.} By introducing the metric duals $p \equiv
\widetilde{P} \equiv g(P,-)$ and $v \equiv \widetilde{V}$ and
denoting interior operator (contraction with any vector $V$) on
forms by $\i_V$ \cite{benn_tucker}, the MPD equations \cite{dixon}
can be written in a compact form as
\begin{eqnarray}
\nabla_V\,{p} &=& \i_V\,\ff
%\nonumber
\label{force}
%\\[10pt]
\end{eqnarray}
\begin{eqnarray}
\nabla_V\,{s} &=& 2\,p \wedge v
%\nonumber
\label{torque}
%\\[10pt]
\end{eqnarray}
with the spin condition \cite{Tulczyjew}
\begin{eqnarray}
\i_P\,s &=& 0
%\nonumber
\label{spin_cnd}
%\\[10pt]
\end{eqnarray}
and forcing term
\begin{eqnarray}
\ff &=& \frac{1}{4}\, \invstar(R_{ab} \wedge \star\,s) \, e^a
\wedge e^b
%\nonumber
\label{curvspin}
\end{eqnarray}%}
in terms of the exterior product $\wedge$ and Hodge map $\star$
associated with a metric with the signature $(-,+,+,+)$. Thus the
``inverse Hodge map'' denoted by $\invstar$ \cite{benn_tucker}
acts on any $q$-form $\omega$ according to $\invstar\omega =
-(-1)^{q(4-q)} \star\omega$ satisfying $ \star (\invstar \omega) =
\invstar (\star \omega) = \omega$. The curvature 2-forms $
R{}^{a}{}_{b}$ in any basis $\{X_a\}$ with co-basis $\{e^a\}$ are
related to the curvature tensor components $R{}^a{}_{bcd}$ by $
2\,R{}^{a}{}_{b}(X_c,X_d) = R{}^a{}_{bcd}$, $(a, b, c, d = 0, 1,
2, 3)$. In terms of the components $\ff_a = \ff(X_a)$, $v^a =
e^a(V)$,  $p^a = e^a(P)$, $s_{ab} = s (X_a,X_b)$, equation
(\ref{curvspin}) becomes
\begin{eqnarray}
\ff_a = -\frac{1}{2}\,R_a{}_{bcd}\,v^b\,s^{cd}. \nonumber
%\label{force1}
%\\[10pt]
\end{eqnarray}
It follows that \cite{tod} the velocity of the spinning test
particle is proportional to the tangent vector $V$ whose
components take the explicit form
\begin{eqnarray}
v^a &=& \frac{ p^c\,v_c }{ p^f\,p_f} \left( p^a
-
\frac{2\,R{}_{bcde}\,p{}^{c}\,s^{ab}\,s{}^{de}} {4\,p^c\,p_c -
R{}_{bcde}\,s{}^{bc}\,s{}^{de}} \right).
%\nonumber
\label{v}
%\\[10pt]
\end{eqnarray}
The freedom to normalise $V$ ensures that $ p^c\,v_c$ is arbitrary
and once a parameterisation of the world-line has been chosen (\ref{v})
permits the computation of the world-line \cite{ehlers}.
The norm of the momentum vector given by
\begin{eqnarray}
{m} &=& \sqrt{-{g}(P,P)}
%\nonumber
%\label{}
%\\[10pt]
\end{eqnarray}
may be identified as the mass of the particle. This is used to
define the normalised momentum vector given by
\begin{eqnarray}
U &=& \frac{P}{m}.
%\nonumber
\label{defU}
%\\[10pt]
\end{eqnarray}
In Minkowski spacetime $U$ coincides with the 4-velocity
$V/\sqrt{-g(V,V)}$ but in general these two vectors differ in curved
spacetime. Using the 1-form $u = \widetilde{U}$ the spin 1-form $l$ is
defined by
%\begin{eqnarray}
$l = -\frac{1}{2}\,\star(u \wedge s) $
%\nonumber
%\label{defj}
%\end{eqnarray}
and the spin vector by $L = \widetilde{l}$. The angular-momentum
2-form $s$ may be related back to the spin 1-form by
\begin{eqnarray}
s &=& 2\,\i_U\,\invstar l.
%\nonumber
\label{s2j}
%\\[10pt]
\end{eqnarray}
The norm of the spin vector $L$ denoted
%\begin{eqnarray}
$\ell = \sqrt{{g}(L, L)} $
%\nonumber
%\label{}
%\\[10pt]
%\end{eqnarray}
defines the ``spin'' of the particle.  The reduced spin vector and
1-form are defined by $\Sigma = L/m$ and $\varsigma = l/m$
respectively such that the norm $\alpha = \sqrt{g(\Sigma,\Sigma)}
= \ell/m$ denotes the spin to mass ratio. It follows from
(\ref{force}), (\ref{torque}) and (\ref{spin_cnd}) that $\alpha$
and $m$ are constants of motion and the vectors $U$ and $\Sigma$
are orthogonal, i.e. $g(U,\Sigma)=0$.

A vector field $K$ on spacetime is a Killing vector field if
$${\cal L}_K g=0$$
$${\cal L}_K {\cal F}=0$$
for any Einstein-Maxwell solution $\{g,{\cal F}\}$ satisfying
\begin{equation}\label{ein_eq}
\GG_a + 8\pi\TT_a = 0
\end{equation}
\begin{equation}\label{max_eq}
\d\star\FF = 0
\end{equation}
where
\begin{equation}\label{em_stress}
\TT_a = \frac{1}{2}\, \left(\i_{X_a}\FF\wedge\star\FF -
\i_{X_a}\!\star\FF\wedge\FF\right)
\end{equation}
in terms of the Einstein 3-forms $\GG_a$ in any basis
$\{X_a\}$~\cite{benn_tucker}.

For each such $K$ one may find a constant along the world-line
$\C$. Sufficient constants of the motion   enable one to integrate
the equations of motion in terms of properties of the background
metric. Given $K = k^a\,X_a$  it follows from \eqref{force},
\eqref{torque} and \eqref{spin_cnd} that the quantity
\begin{eqnarray}
C_K &\equiv& \invstar \left\{ \widetilde{K} \wedge \star p +
\frac{1}{4}\,\d\,\widetilde{K} \wedge \star s \right\} \;=\;
k_a\,p^a - \frac{1}{2}\,k_{[a;b]}\,s^{ab}
%\nonumber
\label{Ci}
%\\[10pt]
\end{eqnarray}
is preserved along the world-line of a spinning test
particle~\cite{dixon, mtw}. Its value can therefore be fixed in
terms of particular values of $p$ and $s$ at any event on their
trajectories.

\section{The MPD equations in an Einstein-Maxwell plane-wave background}

The metric for an Einstein-Maxwell plane wave spacetime takes the
form
\begin{eqnarray}
{g} &=& H(\U, \X, \Y)\,\d \U \otimes \d \U - \frac{1}{2}\left( \d
\U \otimes \d \V + \d \V \otimes \d \U \right) + \d \X \otimes \d
\X + \d \Y \otimes \d \Y
%\nonumber
\label{gwave}
%\\[10pt]
\end{eqnarray}
in ``harmonic'' (Kerr-Schild) coordinates $(\T, \X, \Y, \Z)$. It
describes a plane gravitational and electromagnetic wave
travelling in the $\Z$-direction~\cite{penrose, griffiths}, where
$\U = \T - \Z$, $\V = \T + \Z$ and
\begin{eqnarray}
H(\U, \X, \Y) &=& \phi_{\g}(\U)\,(\X^2 - \Y^2) +  \phi_{\e}(\U)\,(\X^2 +
\Y^2).
%\nonumber
%\label{}
%\\[10pt]
\end{eqnarray}
Here $\phi_{\g}(\U)$ is an arbitrary gravitational wave profile whereas
$\phi_{\e}(\U)$ is related to the electromagnetic potential 1-form
\begin{eqnarray}
\AA &=& \AA_1(\U) \, \d\X + \AA_2(\U) \, \d\Y
\end{eqnarray}
with arbitrary waveforms $\AA_1(\U)$ and $\AA_2(\U)$ by
\begin{equation}\label{}
\phi_{\e}(\U) = -2\pi\left(\AA_1'(\U)^2 + \AA_2'(\U)^2\right)
\end{equation}
and $\;'$ denotes differentiation. The electromagnetic field 2-form $\FF=\d\,\AA$.

In the orthonormal co-basis defined by
\begin{eqnarray}
e^0 &=& \frac{ \d \V }{2} + (1 - H(\U, \X, \Y))\,\frac{\d \U }{2}
\nonumber
%\label{}
\\[10pt]
e^1 &=& \d \X, \;\;
e^2  = \d \Y \nonumber
%\label{}
\\[10pt]
e^3 &=& \frac{ \d \V}{2} - (1 + H(\U, \X, \Y))\,\frac{\d \U }{2}
%\nonumber
\label{kerrschild}
%\\[10pt]
\end{eqnarray}
let  $\{ X_a \}$ be the dual basis satisfying $e^a(X_b) =
\delta^a{}_b$. This basis is parallel along the geodesic observer
$\O: \tau \mapsto (\T(\tau)=\tau, \X=0, \Y=0, \Z=0)$, i.e.
$\nabla\,X_a |_\O= 0$, and takes the form $\{X_0 = \pd{\T}, X_1 =
\pd{\X}, X_2 = \pd{\Y}, X_3 = \pd{\Z}\}$ and sets up a ``local
Lorentz frame'' along $\O$. In these coordinates this observer
curve is parameterised by $\O: \tau \mapsto (\U(\tau)=\tau,
\V(\tau)=\tau, \X=0, \Y=0)$.

We are interested in the motion of a spinning test particle with
mass $m$, spin $\ell = \alpha\,m$ initially at ``rest'' relative
to observer $\O$ and excited by a pulse of  plane gravitational
and electromagnetic waves . Thus we parameterise the world-line of
this particle by $\C: \lambda \mapsto (\U(\lambda)=\lambda,
\V(\lambda), \X(\lambda), \Y(\lambda))$ for functions
$\V(\lambda), \X(\lambda), \Y(\lambda)$ subject to the initial
conditions:
\begin{eqnarray}
\V(0) &=& 0,  \qquad
\X(0) = x_{\in}, \qquad \Y(0) = y_{\in}
%\nonumber
\label{init_VXY}
%\\[10pt]
\end{eqnarray}
with arbitrary position parameters $x_{\in}$ and $y_{\in}$. The
tangent vector $V$ to the worldline $\C$ is written in these
coordinates:
$$V=C^\prime=\partial_{\U} +
\V'(\lambda)\partial_{\V}+\X'(\lambda)\partial_{\X}+\Y'(\lambda)\partial_{\Y}.$$
The pulse nature of the wave can be accommodated by taking the
arbitrary profiles in the Einstein-Maxwell solution above to have
compact support in the variable $\U$.  Where such profiles vanish
the spacetime is flat.

Since the mass $m$ and spin $\alpha$ of the test particles are
constants of the motion one has
\begin{eqnarray}
\eta_{ab}\,u^a(\lambda)\,u^b(\lambda) &=& -1
%\nonumber
\label{p_cnd}
\\[10pt]
\eta_{ab}\,\varsigma^a(\lambda)\,\varsigma^b(\lambda) &=& \alpha^2
%\nonumber
\label{s_cnd}
\\[10pt]
\eta_{ab}\,u^a(\lambda)\,\varsigma^b(\lambda) &=& 0
%\nonumber
\label{ps_cnd}
%\\[10pt]
\end{eqnarray}
in terms of the orthonormal components  $u^a = e^a(U)$ and $
\varsigma^a = e^a( \Sigma)$  with $\eta_{ab} = {\rm
diag}(-1,1,1,1)$. The initial conditions for $u^a$ are:
\begin{eqnarray}
u^0(0) &=& 1, \;
u^1(0) = u^2(0) =  u^3(0) = 0
%\nonumber
\label{init_u}
%\\[10pt]
\end{eqnarray}
and we write the initial conditions for $ \varsigma^a$
\begin{eqnarray}
\varsigma^0(0) &=& 0,\;
\varsigma^1(0) = \varsigma_{\in}^1,\;
\varsigma^2(0) = \varsigma_{\in}^2,\;
\varsigma^3(0) = \varsigma_{\in}^3
%\nonumber
\label{init_s}
%\\[10pt]
\end{eqnarray}
where the constants $\varsigma_{\in}^1$, $\varsigma_{\in}^2$,
$\varsigma_{\in}^3$ satisfy $(\varsigma_{\in}^1)^2 +
(\varsigma_{\in}^2)^2 + (\varsigma_{\in}^3)^2 = \alpha^2$.

For convenience let
\begin{eqnarray}
\phi_\pm(\U) &\equiv& \phi_{\g}(\U) \pm  \phi_{\e}(\U)
%\nonumber
\label{def_f}
\\[10pt]
\Phi_\pm(\U) &\equiv& \int_0^{\U} \phi_\pm(\zeta)\, \d\zeta.
%\nonumber
\label{def_F}
%\\[10pt]
\end{eqnarray}

The Killing symmetry of the Einstein-Maxwell plane wave spacetime
is most conveniently expressed in ``group'' (Rosen) coordinates
$(\u, \v, x, y)$. For suitable ranges these are related to the
previous chart by the relations:
\begin{eqnarray}
\U &=& \u
%\nonumber
\label{rosen_u}
\\[10pt]
\V &=& \v + x^2\,a({\u})\,a'({\u}) + y^2\,b({\u})\,b'({\u})
%\nonumber
%\label{}
\\[10pt]
\X &=& a(\u)\,x
%\nonumber
%\label{}
\\[10pt]
\Y &=& b(\u)\,y
%\nonumber
\label{rosen_y}
%\\[10pt]
\end{eqnarray}
where the metric functions $a(\u)$ and $b(\u)$ satisfy
\begin{eqnarray}
a''(\u) &=& \phi_+(\u)\, a(\u)
%\nonumber
\label{eq_a}
\\[10pt]
b''(\u) &=& -\phi_-(\u)\, b(\u)
%\nonumber
\label{eq_b}
%\\[10pt]
\end{eqnarray}
with a
convenient choice of the (gauge fixing) conditions:
\begin{eqnarray}
a(0) &=& b(0) \;=\; 1
%\nonumber
\label{init_ab}
\\[10pt]
a'(0) &=& b'(0) \;=\; 0.
%\nonumber
\label{init_dfab}
%\\[10pt]
\end{eqnarray}
In this coordinate chart
the metric $g$ takes the form
\begin{eqnarray}
{g} &=& -\frac{1}{2}\left( \d \u \otimes \d \v + \d \v \otimes \d
\u \right) + a(\u)^2\,\d x \otimes \d x + b(\u)^2\,\d {y} \otimes
\d {y}.
%\nonumber
\label{rosen}
%\\[10pt]
\end{eqnarray}

%WHAT HAPPENS AT THE ZEROES OF a and b???

This Einstein-Maxwell spacetime admits five independent Killing
vectors \cite{griffiths}: %IS THIS THE RIGHT REF HERE?
\begin{eqnarray}
K_1 &=& \frac{\partial}{\partial {\v}},\;
K_2 = \frac{\partial}{\partial {x}},\;
K_3 = \frac{\partial}{\partial {y}}
%\nonumber
\label{K1}
\\[10pt]
K_4 &=& A({\u})\,\frac{\partial}{\partial {x}} +
2\,{x}\frac{\partial}{\partial {\v}}
%\nonumber
%\label{}
\\[10pt]
K_5 &=& B({\u})\,\frac{\partial}{\partial {y}} +
2\,{y}\frac{\partial}{\partial {\v}}
%\nonumber
\label{K5}
%\\[10pt]
\end{eqnarray}
where
\begin{eqnarray}
A({\u}) &=& \int_0^\u \frac{ \d \zeta}{a({\zeta})^2} ,\quad
B({\u}) = \int_0^\u \frac{ \d \zeta}{b({\zeta})^2}.
%\nonumber
%\label{}
%\\[10pt]
\end{eqnarray}
In these coordinates we write the tangent vector
$$V=\partial_\u+ \v^\prime(\lambda)\partial _\v +x^\prime(\lambda)\partial _x +y^\prime(\lambda)\partial
_y.$$

%£

It follows from (\ref{Ci}), (\ref{init_VXY}),
(\ref{init_u}) and (\ref{init_s})
that
\begin{eqnarray}
C_{K_1} &=& -\frac{{m}}{2}, \; C_{K_2} = 0, \; C_{K_3} = 0, \;
C_{K_4} = {{m}\beta_2}, \; C_{K_5} = -{{m}\beta_1}
%\nonumber
\label{cc}
%\\[10pt]
\end{eqnarray}
where $\beta_1 = \varsigma_{\in}^1 + y_{\in}$ and $\beta_2 = \varsigma_{\in}^2 - x_{\in}$.
From (\ref{init_VXY}) and (\ref{rosen_u}) -- (\ref{rosen_y})
in group coordinates
the initial conditions for
the world-line of the spinning test particle
$\C: \lambda \mapsto (\u(\lambda)=\lambda, \v(\lambda),
x(\lambda), y(\lambda))$ read
\begin{eqnarray}
\v(0) &=& 0, \qquad
x(0) = x_{\in}, \qquad y(0) = y_{\in}.
%\nonumber
\label{init_vxy}
%\\[10pt]
\end{eqnarray}

%CHECK TEX LABELS  for 20,21,22,42,43 IN TEX FILE:

The five Killing vectors above yield five algebraic
relations, \eqref{Ci}, between the components of $u$ and $\varsigma$ for
any $\lambda$. Equations \eqref{p_cnd}, \eqref{s_cnd}, \eqref{ps_cnd}
provide three further
algebraic relations. These eight equations may be solved in terms of
the constants of motion and initial conditions to yield the
solutions \eqref{pp} and \eqref{ss} below:
\begin{eqnarray}
u^0(\lambda)
&=&
1
+
\frac{\beta_2^2}{2} \,a'(\lambda)^2 + \frac{\beta_1^2}{2}\,b'(\lambda)^2
\nonumber
%\label{pp0}
\\[10pt]
u^1(\lambda) &=& -\beta_2 a'(\lambda),\quad
u^2(\lambda)  =  \beta_1 b'(\lambda)
\nonumber
%\label{pp2}
\\[10pt]
u^3(\lambda)
&=&
%{\beta_2^2 \over 2} \,a'(\lambda)^2 + {\beta_1^2 \over 2}\,b'(\lambda)^2
\frac{\beta_2^2}{2} \,a'(\lambda)^2 + \frac{\beta_1^2}{2}\,b'(\lambda)^2
%\nonumber
\label{pp}
%\\[10pt]
\end{eqnarray}
%while
%(\ref{ps_cnd}) and (\ref{cc}) give
and
\begin{eqnarray}
\varsigma^0(\lambda)
&=&
\frac{1}{
\beta_2^2\,a'(\lambda)^2
+
\beta_1^2\,b'(\lambda)^2
-2
}
\left\{
\left(
\beta_2^2\,a'(\lambda)^2
+
\beta_1^2\,b'(\lambda)^2
\right)\varsigma^3(\lambda)
\right.
\nonumber
%\label{}
\\[10pt]
&&
-
\left. 2 (y(\lambda)-\beta_1) \beta_2 b(\lambda) a'(\lambda)
-
2 (x(\lambda)+\beta_2) \beta_1 a(\lambda) b'(\lambda) \right\}
\nonumber
%\label{ss0}
\\[10pt]
\varsigma^1(\lambda) &=& -(y(\lambda)-\beta_1) b(\lambda) +
\frac{1}{ \beta_2^2\,a'(\lambda)^2 + \beta_1^2\,b'(\lambda)^2 -2
} \left\{ -2 \beta_2{}\,a'(\lambda) \varsigma^3(\lambda) \right.
\nonumber
%\label{}
\\[10pt]
&& \left. +2\,{\beta_2{}} \,a'(\lambda) \left[
(y(\lambda)-\beta_1) \beta_2 b(\lambda) a'(\lambda) +
(x(\lambda)+\beta_2) \beta_1 a(\lambda) b'(\lambda) \right]
\right\}
\nonumber
%\label{ss1}
\\[10pt]
\varsigma^2(\lambda) &=& (x(\lambda)+\beta_2) a(\lambda) +
\frac{1}{ \beta_2^2\,a'(\lambda)^2 + \beta_1^2\,b'(\lambda)^2 -2
} \left\{ 2 \beta_1{}\,b'(\lambda) \varsigma^3(\lambda) \right.
\nonumber
%\label{}
\\[10pt]
&& \left. -2\,{\beta_1{}} \,b'(\lambda) \left[
(y(\lambda)-\beta_1) \beta_2 b(\lambda) a'(\lambda) +
(x(\lambda)+\beta_2) \beta_1 a(\lambda) b'(\lambda) \right]
\right\}.
%\nonumber
\label{ss}
%\\[10pt]
\end{eqnarray}

Furthermore they  imply a quadratic equation for
$\varsigma^3(\lambda)$:
\begin{eqnarray}
\varsigma^3(\lambda)^2 + {\cal P}(\lambda)\,\varsigma^3(\lambda)
+ {\cal Q}(\lambda)
&=&
0
%\nonumber
\label{ss3}
%\\[10pt]
\end{eqnarray}
where
\begin{eqnarray}
{\cal P}(\lambda) &=& -2\,(y(\lambda)-\beta_1)
\beta_2 b(\lambda) a'(\lambda)
-2\,(x(\lambda)+\beta_2) \beta_1 a(\lambda)
b'(\lambda)
%\nonumber
%\label{ss0}
%\\[10pt]
\end{eqnarray}
and
\begin{eqnarray}
{\cal Q}(\lambda)
&=&
-\frac{\beta^2}{4}
\left\{
\beta_2^2\,a'(\lambda)^2
+
\beta_1^2\,b'(\lambda)^2
-2
\right\}^2
\nonumber
%\label{ss0}
\\[10pt]
&& \hspace{-20pt} + \frac{(y(\lambda)-\beta_1)^2
b(\lambda)^2}{4} \left\{ 2 \beta_1^2 \beta_2^2 a'(\lambda)^2
b'(\lambda)^2 + \beta_2^4 a'(\lambda)^4 + \beta_1^4
b'(\lambda)^4
-
4 \beta_1^2 b'(\lambda)^2
+
4
\right\}
\nonumber
%\label{ss0}
\\[10pt]
&& \hspace{-20pt} + \frac{(x(\lambda)+\beta_2)^2
a(\lambda)^2}{4} \left\{ 2 \beta_1^2 \beta_2^2 a'(\lambda)^2
b'(\lambda)^2 + \beta_2^4 a'(\lambda)^4 + \beta_1^4
b'(\lambda)^4
-
4 \beta_2^2 a'(\lambda)^2
+
4
\right\}
\nonumber
%\label{}
\\[10pt]
&& \hspace{-20pt} + {2}\, \beta_1 \beta_2 a(\lambda) b(\lambda)
a'(\lambda) b'(\lambda) (x(\lambda)+\beta_2)
(y(\lambda)-\beta_1).
%\nonumber
%\label{}
%\\[10pt]
\end{eqnarray}
Thus for given metric functions $a(\lambda),\,b(\lambda)$ one has
solutions $\varsigma^3(\lambda)$ in terms of $x(\lambda)$ and
$y(\lambda)$.

The differential equations for the components $\v(\lambda),
x(\lambda), y(\lambda)$  of the tangent vector $V$ are obtained
from substituting (\ref{defU}), (\ref{s2j}), (\ref{pp}) and
(\ref{ss}) into (\ref{v}):
\begin{eqnarray}
x'(\lambda) &=& 2\,\phi_-(\lambda) \left\{ \beta_1 a(\lambda)
b(\lambda) b'(\lambda) (x(\lambda)+\beta_2) (y(\lambda)-\beta_1) +
\beta_2 b(\lambda)^2 a'(\lambda) (y(\lambda)-\beta_1)^2 \right.
\nonumber
%\label{}
\\[10pt]
&&
%\hspace{-20pt}
\left. - (y(\lambda)-\beta_1) b(\lambda)
\varsigma^3(\lambda) \right\}/\left\{ a(\lambda) \left(
\beta_2^2\,a'(\lambda)^2 + \beta_1^2\,b'(\lambda)^2 -2 \right)
\right\} -\frac{ (x(\lambda)+\beta_2) a'(\lambda)}{a(\lambda)}
%\nonumber
\label{dx}
%\\[10pt]
\end{eqnarray}
\begin{eqnarray}
y'(\lambda) &=& 2\,\phi_+(\lambda) \left\{ \beta_2 a(\lambda)
b(\lambda) a'(\lambda) (x(\lambda)+\beta_2) (y(\lambda)-\beta_1) +
\beta_1 a(\lambda)^2 b'(\lambda) (x(\lambda)+\beta_2)^2 \right.
\nonumber
%\label{}
\\[10pt]
&&
%\hspace{-20pt}
\left.
-
(x(\lambda)+\beta_2) a(\lambda) \varsigma^3(\lambda) \right\}
/\left\{ b(\lambda) \left( \beta_2^2\,a'(\lambda)^2 +
\beta_1^2\,b'(\lambda)^2 -2 \right) \right\} -\frac{
(y(\lambda)-\beta_2) b'(\lambda)}{b(\lambda)}
%\nonumber
\label{dy}
%\\[10pt]
\end{eqnarray}
\begin{eqnarray}
\v'(\lambda)
&=&
1
+ \left( x(\lambda)+\beta_{{2}} \right) ^{2} a'(\lambda) ^{2}
+\left( y(\lambda) -\beta_{{1}} \right) ^{2} b'(\lambda) ^{2}
\nonumber\\[10pt]
&&
\hspace{-25pt}
+\frac{\phi_+(\lambda)}{\beta_2^2 a'(\lambda) ^{2}+\beta_1^2 b'(\lambda) ^{2}-2}
\left\{ -2\,a(\lambda)^2\left( x(\lambda)+\beta_{{2}}  \right)
^{2} \left( \beta_2^2 a'(\lambda) ^{2}
+2\,\beta_{{1}}\, b'(\lambda) ^{2} y(\lambda) -\beta_1^2 b'(\lambda) ^{2}-2 \right)
\right.
\nonumber\\[10pt]
&&
\hspace{-25pt}
\left.
-4\, \beta_{{2}}\, a(\lambda) b(\lambda) a'(\lambda)  b'(\lambda)
\left( x(\lambda)+\beta_{{2}}
 \right)  \left( y(\lambda) -\beta_{{1}} \right) ^{2}
 +4\,a(\lambda)  b'(\lambda) \varsigma^3(\lambda) \left( x(\lambda)+\beta_{{2}}  \right)
 \left( y(\lambda) -\beta_{{1}} \right)
\right\}
\nonumber\\[10pt]
&&
\hspace{-25pt}
+\frac{\phi_-(\lambda)}{\beta_2^2 a'(\lambda) ^{2}+\beta_1^2 b'(\lambda) ^{2}-2}
\left\{
2\, b(\lambda)^2\left( y(\lambda) -\beta_{{1}}
 \right) ^{2} \left( \beta_1^2 b'(\lambda) ^{2}
 -2\,\beta_{{2}} \,a'(\lambda)^2 x(\lambda) -\beta_2^2 a'(\lambda) ^{2}-2 \right)
\right.
\nonumber\\[10pt]
&&
\hspace{-25pt}
\left.
-4\,\beta_{{1}}\, a(\lambda) b(\lambda) a'(\lambda) b'(\lambda)
\left( x(\lambda)+\beta_{{2}}
 \right) ^{2} \left( y(\lambda) -\beta_{{1}} \right)
+4\,  b(\lambda)  a'(\lambda)  \varsigma^3(\lambda) \left( y(\lambda) -\beta_{{1}}
 \right)
 \left( x(\lambda)+\beta_{{2}}  \right)\right\}
%
%1 +
%{(x(\lambda)+\beta_2)^2 a'(\lambda)^2} +
%{(y(\lambda)-\beta_1)^2 b'(\lambda)^2}
%\nonumber
%\label{}
%\\[10pt]
%&& +{\phi_+(\lambda)}\left\{ 4 \varsigma^3(\lambda)
%(x(\lambda)+\beta_2) (y(\lambda)-\beta_1) (a(\lambda) b'(\lambda)
%+ b(\lambda) a'(\lambda)) \right. \nonumber
%\label{}
%\\[10pt]
%&& \left. +2\,(x(\lambda)+\beta_2)^2 a(\lambda)^2 \left( 2-
%\beta_2^2\,a'(\lambda)^2 + \beta_1^2\,b'(\lambda)^2
%-
%2 \beta_1\,b'(\lambda)^2 y(\lambda) \right) \right. \nonumber
%\label{}
%\\[10pt]
%&& \left. -2\,(y(\lambda)-\beta_1)^2 b(\lambda)^2 \left( 2+
%\beta_2^2\,a'(\lambda)^2
%-
%\beta_1^2\,b'(\lambda)^2 + 2 \beta_2\,a'(\lambda)^2 x(\lambda)
%\right) \right. \nonumber
%\label{}
%\\[10pt]
%&& \left. -4\, a(\lambda) b(\lambda) a'(\lambda) b'(\lambda)
%(x(\lambda)+\beta_2) (y(\lambda)-\beta_1) (\beta_1
%x(\lambda) + \beta_2 y(\lambda)) \right\} \nonumber
%\label{}
%\\[10pt]
%&&
%/\left(
%\beta_2^2\,a'(\lambda)^2
%+
%\beta_1^2\,b'(\lambda)^2
%-2
%\right).
%\nonumber
%\\[10pt]
\label{dv}
\end{eqnarray}

These equations are decoupled as the sets:

\begin{eqnarray}
\left(\begin{array}{cc}
    x(\lambda) \\
    y(\lambda)
  \end{array}\right)^\prime
  =
 \left(\begin{array}{cc}
    F_1(x(\lambda),y(\lambda), \varsigma^3(\lambda))   \\
    F_2(x(\lambda),y(\lambda), \varsigma^3(\lambda))
  \end{array}\right)\label{popxy}
 \end{eqnarray}

and

\begin{eqnarray}\v^\prime(\lambda)=F_3(x(\lambda),y(\lambda), \varsigma^3(\lambda)
)\label{popv}\end{eqnarray}

where from (\ref{ss3}),
$$\varsigma^3(\lambda)=F_4(x(\lambda),y(\lambda)).$$

Thus the world-line of the spinning particle can be determined by
solving the two coupled non-linear differential equations
(\ref{popxy})  for $x(\lambda)$ and $y(\lambda)$ and inserting
these solutions into (\ref{popv}) and solving for $\v(\lambda)$.
The evolution of the reduced spin vector $\Sigma$ then follows by
solving (\ref{ss}) and (\ref{ss3}).

Given the initial position parameters $x_{\in}, y_{\in}$ and reduced
spin vector components $\varsigma_{\in}^1$, $\varsigma_{\in}^2$,
$\varsigma_{\in}^3$ the evolution of the normalised momentum vector
$U$ follows more simply from  (\ref{pp}) for arbitrary profiles
$\phi_\pm(\u)$ and the corresponding $a(\u)$, $b(\u)$ obtained
from (\ref{eq_a}), (\ref{eq_b}), (\ref{init_ab}),
(\ref{init_dfab}).

\section{Particular solutions describing geodesic and non-geodesic motions}

Since the above equations are  non-linear it is difficult to find
general analytic solutions for the particle's worldline and  spin.
It is however possible to find interesting particular analytic
solutions for special initial conditions. Thus if  the initial
spin vector is in the direction of propagation of the
gravitational and/or electromagnetic wave then it remains so and
furthermore the subsequent motion is geodesic. With
$\varsigma_{\in}^1=\varsigma_{\in}^2=0, \alpha = |\varsigma_{\in}^3|$
and any $x_{\in}$ and $y_{\in}$, the world-line $\C$ is simply:
\begin{equation}\label{}
  \u(\lambda) = \lambda,\;
  x(\lambda) = x_{\in},\;
  y(\lambda) = y_{\in},\;
  \v(\lambda) = \lambda
\end{equation}
with normalised momentum  components:
\begin{eqnarray}
u^0(\lambda{})
&=&
1 + \frac{x_{\in}^2 }{ 2} \, a'(\lambda)^2 +
\frac{y_{\in}^2}{2}
b'(\lambda{})^2
\nonumber
\\[10pt]
u^1(\lambda{}) &=& x_{\in} \, a'(\lambda),\;\;
u^2(\lambda{}) =
y_{\in}\, b'(\lambda{})
\nonumber
\\[10pt]
u^3(\lambda{}) &=& \frac{x_{\in}^2 }{ 2} \, a'(\lambda)^2 +
\frac{y_{\in}^2 }{2}\,b'(\lambda{})^2
%\nonumber
\label{sol_u3}
%\\[10pt]
\end{eqnarray}
and reduced spin  components:
\begin{eqnarray}
\varsigma^0(\lambda{})
&=&
-\frac{\varsigma_{\in}^3}{2}\left(
x_{\in}^2 \, a'(\lambda)^2 + y_{\in}^2 \, b'(\lambda{})^2\right)
\nonumber
\\[10pt]
\varsigma^1(\lambda{}) &=&  -\varsigma_{\in}^3\,x_{\in} \, a'(\lambda),\quad
\varsigma^2(\lambda{}) = -\varsigma_{\in}^3\,y_{\in}\, b'(\lambda{})
\nonumber
\\[10pt]
\varsigma^3(\lambda{}) &=&
\frac{\varsigma_{\in}^3}{2}\left(
1-
x_{\in}^2 \, a'(\lambda)^2 - y_{\in}^2 \, b'(\lambda{})^2\right).
%\nonumber
\label{sol_s3}
%\\[10pt]
\end{eqnarray}

A class of solutions describing {\em non-geodesic} motions can
also be found in which the particle spin vector is transported
$\nabla$-parallel along its world-line, transverse to the
direction of wave propagation. Consider the case with $\alpha =
|\varsigma_{\in}^1|$, $\varsigma_{\in}^2 = \varsigma_{\in}^3 = 0$ and
\begin{eqnarray}
x(\lambda) &=& x_{\in}.
%\nonumber
\label{x_case1}
%\\[10pt]
\end{eqnarray}
Then (\ref{dx}) implies
\begin{eqnarray}
\varsigma^3(\lambda) = x_{\in} \left(\varsigma_{\in}^1 + y_{\in} - y(\lambda)
\right) b(\lambda) a'(\lambda)
%\nonumber
\label{}
%\\[10pt]
\end{eqnarray}
and (\ref{dy}) becomes
\begin{eqnarray}
y'(\lambda)
&=&
\frac{(\varsigma_{\in}^1 + y_{\in}  -
y(\lambda)) b'(\lambda)}{b(\lambda)}
%\nonumber
%\label{}
%\\[10pt]
\end{eqnarray}
with the solution
\begin{eqnarray}
y(\lambda) &=& {\varsigma_{\in}^1} \left(1 -
\frac{1}{b(\lambda)}\right) +y_{\in}
%\nonumber
\label{y_case1}
%\\[10pt]
\end{eqnarray}
satisfying the initial condition $y(0)=y_{\in}$. Equation (\ref{dv})
then reduces to
\begin{eqnarray}
\v'(\lambda) &=& 1 + {(\varsigma_{\in}^1)^2}
\left(2\,\phi_-(\lambda{}) + \frac{b'(\lambda)^2}{b(\lambda)^2}
\right).
\end{eqnarray}
Using (\ref{eq_b}) and the initial condition $\v(0) =0$ this can
be integrated :
\begin{eqnarray}
\v(\lambda) &=& \lambda + {(\varsigma_{\in}^1)^2}\left( \Phi_-(\lambda)
- \frac{b'(\lambda)}{b(\lambda)} \right).
%\nonumber
\label{v_case1}
%\\[10pt]
\end{eqnarray}
Substituting the above equations into (\ref{pp}) and (\ref{ss})
gives
\begin{eqnarray}
u^0(\lambda{})
&=&
1 + \frac{x_{\in}^2 }{ 2} \, a'(\lambda)^2 +
\frac{(\varsigma_{\in}^1+y_{\in})^2}{2}
b'(\lambda{})^2
\nonumber
\\[10pt]
u^1(\lambda{}) &=& x_{\in} \, a'(\lambda),\quad
u^2(\lambda{}) =
(\varsigma_{\in}^1+y_{\in})\, b'(\lambda{})
\nonumber
\\[10pt]
u^3(\lambda{}) &=& \frac{x_{\in}^2 }{ 2} \, a'(\lambda)^2 +
\frac{(\varsigma_{\in}^1+y_{\in})^2 }{2}\,b'(\lambda{})^2
%\nonumber
\label{sol_u1}
%\\[10pt]
\end{eqnarray}
and
\begin{eqnarray}
\varsigma^0(\lambda{}) &=& \varsigma_{\in}^1 \, x_{\in} \, a'(\lambda),\quad
\varsigma^1(\lambda{}) = \varsigma_{\in}^1 ,\quad
\varsigma^2(\lambda{}) = 0 ,\quad
\varsigma^3(\lambda{}) = \varsigma_{\in}^1 \, x_{\in} \, a'(\lambda).
%\nonumber
%\label{}
%\\[10pt]
\end{eqnarray}
Using (\ref{rosen_u}) --  (\ref{rosen_y}) solutions
(\ref{x_case1}), (\ref{y_case1}), (\ref{v_case1}) in harmonic
coordinates become
\begin{eqnarray}
\T(\lambda)
&=&
\lambda +
\frac{x_{\in}^2 }{ 2} \, a(\lambda) \, a'(\lambda) +
\frac{(\varsigma_{\in}^1+y_{\in})^2 }{ 2} \, b(\lambda) \, b'(\lambda)
%\nonumber
%\\[10pt]
%& &{}
- 2 \, \varsigma_{\in}^1(\varsigma_{\in}^1+y_{\in}) \, b'(\lambda) +
\frac{(\varsigma_{\in}^1)^2 }{ 2} \, \Phi_-(\lambda) \nonumber
%\label{sol_t1}
\\[10pt]
\X(\lambda) &=& a(\lambda) \, x_{\in},\quad
\Y(\lambda)  =  b(\lambda)(\varsigma_{\in}^1+y_{\in}) - \varsigma_{\in}^1
\nonumber
%\label{sol_y1}
\\[10pt]
\Z(\lambda) &=&
\frac{x_{\in}^2}{ 2} \, a(\lambda) \, a'(\lambda) +
\frac{(\varsigma_{\in}^1+y_{\in})^2 }{ 2} \, b(\lambda) \, b'(\lambda)
%\nonumber
%\\[10pt]
%& &{}
- 2 \, \varsigma_{\in}^1(\varsigma_{\in}^1+y_{\in}) \, b'(\lambda) +
\frac{(\varsigma_{\in}^1)^2 }{ 2} \, \Phi_-(\lambda).
%\nonumber
\label{sol_1}
%\\[10pt]
\end{eqnarray}
For such non-geodesic motion  the difference between the
corresponding  momentum and velocity vectors is nonzero and takes
the form:
\begin{eqnarray}
U -  V & = & -2 \, (\varsigma_{\in}^1)^2 \phi_-(\lambda)
\frac{\partial }{
\partial \V}.
\end{eqnarray}

With the alternative initial spin conditions $\varsigma_{\in}^1
=\varsigma_{\in}^3 = 0$ and $\alpha = |\varsigma_{\in}^2|$, one
similarly finds:
\begin{eqnarray}
u^0(\lambda{})
&=&
1 + \frac{(\varsigma_{\in}^2 - x_{\in})^2}{2}
a'(\lambda{})^2 + \frac{y_{\in}^2 }{ 2} \, b'(\lambda)^2,
\nonumber
\\[10pt]
u^1(\lambda{}) &=& -(\varsigma_{\in}^2 - x_{\in}) \,a'(\lambda{}) ,\quad
u^2(\lambda{})  = y_{\in} \, b'(\lambda)
\nonumber
\\[10pt]
u^3(\lambda{}) &=& \frac{(\varsigma_{\in}^2 - x_{\in})^2 }{2}\,a'(\lambda{})^2
+ \frac{y_{\in}^2 }{ 2} \, b'(\lambda)^2
%\nonumber
\label{sol_u2}
%\\[10pt]
\end{eqnarray}
and
\begin{eqnarray}
\varsigma^0(\lambda{}) &=& \varsigma_{\in}^2 \, y_{\in} \,  b'(\lambda),\quad
\varsigma^1(\lambda{}) = 0 ,\quad
\varsigma^2(\lambda{}) = \varsigma_{\in}^2,\quad
\varsigma^3(\lambda{}) = \varsigma_{\in}^2 \, y_{\in} \, b'(\lambda).
%\nonumber
%\label{}
%\\[10pt]
\end{eqnarray}
The corresponding world-line is then given by
\begin{eqnarray}
\T(\lambda) &=& \lambda \frac{(\varsigma_{\in}^2 - x_{\in})^2 }{ 2} \,
a(\lambda) \, a'(\lambda) + \frac{y_{\in}^2 }{ 2} \, b(\lambda) \,
b'(\lambda) - 2 \, \varsigma_{\in}^2(\varsigma_{\in}^2 - x_{\in}) \,
a'(\lambda) - \frac{(\varsigma_{\in}^2)^2 }{ 2} \, \Phi_+(\lambda)
%\nonumber
\label{sol_t2}
\\[10pt]
\X(\lambda) &=& -a(\lambda)(\varsigma_{\in}^2 - x_{\in}) + \varsigma_{\in}^2
%\nonumber
\label{sol_x2}
\\[10pt]
\Y(\lambda) &=& b(\lambda) \, y_{\in}
%\nonumber
\label{sol_y2}
\\[10pt]
\Z(\lambda) &=&
\frac{(\varsigma_{\in}^2 - x_{\in})^2}{ 2} \, a(\lambda) \, a'(\lambda) +
\frac{y_{\in}^2 }{ 2} \, b(\lambda) \, b'(\lambda)
 - 2 \, \varsigma_{\in}^2(\varsigma_{\in}^2 - x_{\in}) \, a'(\lambda) -
\frac{(\varsigma_{\in}^2)^2 }{ 2} \, \Phi_+(\lambda)
%\nonumber
\label{sol_z2}
%\\[10pt]
\end{eqnarray}
in harmonic coordinates. In this case the difference between
momentum and velocity vectors is:
\begin{eqnarray}
U - V & = & 2 \, (\varsigma_{\in}^2)^2 \phi_{+}(\lambda) \frac{\partial}{
\partial \V}.
\end{eqnarray}

\section{Scattering cross-section due to an Einstein-Maxwell wave pulse}

Aside from regularity conditions the above calculations have not
imposed any strong conditions on the structure of the profile
functions that characterise the Einstein-Maxwell background
spacetime. Consider now the scattering of neutral massive spinning
particles by a plane gravitational and electromagnetic wave pulse
represented by the metric \eqref{gwave} with functions $\phi_{\g}(\U)$
and $\phi_{\e}(\U)$ that vanish  for $\U < 0$ and $\U > \delta$,
where $\delta$ denotes the width of the wave pulse. Such a
sandwich wave gives rise to two half Minkowski spacetimes namely
$\M_{\in}$ for $\U<0$ and $\M_{\out}$ for $\U>\delta$ corresponding to
the domains before and after the passage of the plane wave pulse.
Within each half Minkowski spacetime the orthonormal basis $\{X_0
= \pd{\T}, X_1 = \pd{\X}, X_2 = \pd{\Y}, X_3 = \pd{\Z}\}$ defines
a  Lorentz frame, and $\{ X_a \}$ along the geodesic observer $\O$
still remains a local Lorentz basis also in the ``sandwiched''
domain $\D$ where $ 0 \le \U = \T - \Z \le \delta$. Consider a
collection of spinning  particles, labelled by their initial $\X$
and $\Y$ components, i.e. $x_{\in}$ and $y_{\in}$. Suppose these
particles are initially at rest relative to $\O$ having common
initial velocity and momentum given by $V(0) = U(0) = X_0$ but
with possibly non-uniform initial spin orientations described by:
\begin{align}\label{alpha_xy}
  \varsigma_{\in}^1 = \varsigma_{\in}^1(x_{\in},y_{\in}),\quad
  \varsigma_{\in}^2 = \varsigma_{\in}^2(x_{\in},y_{\in}),\quad
  \varsigma_{\in}^3 = \varsigma_{\in}^3(x_{\in},y_{\in}).
\end{align}
Particles scattered by the  pulse will have different out-going
momentum vectors, depending on their initial data. In $\M_{\in}$ and
$\M_{\out}$ these vectors have different constant components $u^a$
and changes in these quantities along $\C$ across the domain $\D$
reflect their gain in energy-momentum  in the Lorentz frame
associated with $\O$. In any chart for $\M_{\out}$ write $a'_{\out}
\equiv a'(\lambda)$, $b'_{\out} \equiv b'(\lambda)$ and $p^a_{\out}
\equiv m\,u^a(\lambda)$ for any $\lambda > \delta$. In such a
sandwich spacetime the components of the final momentum vector are
readily evaluated from (\ref{pp}):
\begin{eqnarray}
p^0_{\out}
&=&
m
+
\frac{m }{ 2}\,(\varsigma_{\in}^2 - x_{\in})^2 \,a'_{\out}{}^2
+ \frac{m }{ 2}\,(\varsigma_{\in}^1 + y_{\in})^2\,b'_{\out}{}^2
%\nonumber
\label{ppp0}
\\[10pt]
p^1_{\out} &=& -m\,(\varsigma_{\in}^2 - x_{\in})\, a'_{\out}
%\nonumber
\label{ppp1}
\\[10pt]
p^2_{\out} &=& m\,(\varsigma_{\in}^1 + y_{\in})\, b'_{\out}
%\nonumber
\label{ppp2}
\\[10pt]
p^3_{\out}
&=&
%\frac{1}{2}\, \left\{
\frac{m }{ 2}\,(\varsigma_{\in}^2 - x_{\in})^2 \,a'_{\out}{}^2
+ \frac{m }{ 2}\,(\varsigma_{\in}^1 + y_{\in})^2\,b'_{\out}{}^2.
%\right\}
%\nonumber
\label{ppp3}
%\\[10pt]
\end{eqnarray}

Equations  \eqref{ppp1} and \eqref{ppp2} define a map between the
transverse components of momentum in any spacelike plane
associated with $\O$'s rest space in $\M_{\out}$ and the initial
location of particles in a similar plane in $\M_{\in}$. It is
therefor natural to define a  classical differential scattering
cross-section associated with this map in terms of the out-going
momentum components $(p^1_{\out},p^2_{\out})$:
\begin{eqnarray}
%d\sigma_\cls
d\sigma
\equiv d x_{\in}\, d y_{\in} & = & \frac{d p^1_{\out} d p^2_{\out}
}{ m^2 \, a'_{\out} \, b'_{\out}} \left(1 + \frac{\partial \varsigma_{\in}^1 }{
\partial y_{\in}} - \frac{\partial \varsigma_{\in}^2 }{ \partial x_{\in}} + \frac{\partial
\varsigma_{\in}^1 }{ \partial x_{\in}} \, \frac{\partial \varsigma_{\in}^2 }{ \partial
y_{\in}} - \frac{\partial \varsigma_{\in}^1 }{ \partial y_{\in}} \, \frac{\partial
\varsigma_{\in}^2 }{ \partial x_{\in}} \right)^{-1}. \label{dsigma}
\end{eqnarray}
This cross-section provides a means of estimating the effects of a
non-uniform distribution of spin on the scattering of particles by
a passing Einstein-Maxwell pulse of radiation. When this initial
polarisation is uniform (i.e.  $\varsigma_{\in}^1$,
$\varsigma_{\in}^2$ are independent of the parameters $x_{\in}$,
$y_{\in}$) this expression agrees with the differential
cross-section describing the scattering of classical or quantum
non-spinning particles~\cite{garriga} by an Einstein-Maxwell pulse
and furthermore  coincides with the differential cross-section
describing the scattering of {\em polarised} incident quantum
Dirac particles in such a background~\cite{bini_ferrari}. A priori
there is no reason to expect that electrically neutral particles
with spin-curvature interactions described by the MPD equations
should behave in this way. More generally we believe that the
cross section (\ref{dsigma}) describes how such particles with  a
non-uniform spatial distribution of spins are scattered by an
Einstein-Maxwell pulse in space. In the absence of a relativistic
transport description of astrophysical phenomena involving
spin-curvature interactions the approach above offers a relatively
straightforward mechanism to estimate the significance of such
interactions.

\section{Conclusions}

 By exploiting the
high symmetry of  Einstein-Maxwell plane-wave spacetimes we have
reduced the pole-dipole MPD equations for the motion of massive
particles with spin to a system of tractable ordinary differential
equations. Classes of exact solutions have been found
corresponding to particular initial conditions for the directions
of the particle spin relative to the direction of the propagating
background fields. For  Einstein-Maxwell pulses we have defined a
scattering cross section that reduces in certain limits to those
associated with the scattering of scalar and Dirac particles based
on classical and quantum field theoretic techniques. Such
techniques are considerably more intricate than those discussed
here. The relative simplicity of our approach and its use of
macroscopic spin distributions suggests that it may have
advantages in those astrophysical situations that involve strong
classical gravitational and electromagnetic environments.

\section{Acknowledgements}

The authors are grateful to EPSRC and BAE Systems for financial
support.

%NEED TO WEED OUT ANY REFS THAT ARE NOT REFERRED TO

\end{document}